\documentclass{webofc}
\usepackage[varg]{txfonts}   
\begin{document}
\title{GASTON: Galactic Star Formation with NIKA2 }
\subtitle{A new population of cold massive sources discovered}
%
%

\author{\firstname{N.} \lastname{Peretto} \inst{\ref{Cardiff}}\fnsep\thanks{\email{nicolas.peretto@astro.cf.ac.uk}}
\and \firstname{A.} \lastname{Rigby}\inst{\ref{Cardiff}}
\and \firstname{R.} \lastname{Adam} \inst{\ref{LLR},\ref{CEFCA}}
\and  \firstname{P.} \lastname{Ade} \inst{\ref{Cardiff}}
\and  \firstname{P.} \lastname{Andr\'e} \inst{\ref{CEA1}}
\and  \firstname{A.} \lastname{Andrianasolo} \inst{\ref{IPAG}}
\and  \firstname{H.} \lastname{Aussel} \inst{\ref{CEA1}}
\and  \firstname{A.} \lastname{Bacmann} \inst{\ref{IPAG}}
\and  \firstname{A.} \lastname{Beelen} \inst{\ref{IAS}}
\and  \firstname{A.} \lastname{Beno\^it} \inst{\ref{Neel}}
\and  \firstname{A.} \lastname{Bideaud} \inst{\ref{Neel}}
\and  \firstname{O.} \lastname{Bourrion} \inst{\ref{LPSC}}
\and  \firstname{M.} \lastname{Calvo} \inst{\ref{Neel}}
\and  \firstname{A.} \lastname{Catalano} \inst{\ref{LPSC}}
\and  \firstname{B.} \lastname{Comis} \inst{\ref{LPSC}}
\and  \firstname{M.} \lastname{De~Petris} \inst{\ref{Roma}}
\and  \firstname{F.-X.} \lastname{D\'esert} \inst{\ref{IPAG}}
\and  \firstname{S.} \lastname{Doyle} \inst{\ref{Cardiff}}
\and  \firstname{E.~F.~C.} \lastname{Driessen} \inst{\ref{IRAMF}}
\and  \firstname{A.} \lastname{Gomez} \inst{\ref{CAB}}
\and  \firstname{J.} \lastname{Goupy} \inst{\ref{Neel}}
\and  \firstname{F.} \lastname{K\'eruzor\'e} \inst{\ref{LPSC}}
\and  \firstname{C.} \lastname{Kramer} \inst{\ref{IRAME}}
\and  \firstname{B.} \lastname{Ladjelate} \inst{\ref{IRAME}}
\and  \firstname{G.} \lastname{Lagache} \inst{\ref{LAM}}
\and  \firstname{S.} \lastname{Leclercq} \inst{\ref{IRAMF}}
\and  \firstname{J.-F.} \lastname{Lestrade} \inst{\ref{LERMA}}
\and  \firstname{J.F.} \lastname{Mac\'ias-P\'erez} \inst{\ref{LPSC}}
\and  \firstname{P.} \lastname{Mauskopf} \inst{\ref{Cardiff},\ref{Arizona}}
\and \firstname{F.} \lastname{Mayet} \inst{\ref{LPSC}}
\and  \firstname{A.} \lastname{Monfardini} \inst{\ref{Neel}}
\and  \firstname{F.} \lastname{Motte} \inst{\ref{IPAG}}
\and  \firstname{L.} \lastname{Perotto} \inst{\ref{LPSC}}
\and  \firstname{G.} \lastname{Pisano} \inst{\ref{Cardiff}}
\and  \firstname{N.} \lastname{Ponthieu} \inst{\ref{IPAG}}
\and  \firstname{V.} \lastname{Rev\'eret} \inst{\ref{CEA1}}
\and  \firstname{I.} \lastname{Ristorcelli} \inst{\ref{IRAP}}
\and  \firstname{A.} \lastname{Ritacco} \inst{\ref{IRAME}}
\and  \firstname{C.} \lastname{Romero} \inst{\ref{IRAMF}}
\and  \firstname{H.} \lastname{Roussel} \inst{\ref{IAP}}
\and  \firstname{F.} \lastname{Ruppin} \inst{\ref{MIT}}
\and  \firstname{K.} \lastname{Schuster} \inst{\ref{IRAMF}}
\and  \firstname{S.} \lastname{Shu} \inst{\ref{IRAMF}}
\and  \firstname{A.} \lastname{Sievers} \inst{\ref{IRAME}}
\and  \firstname{C.} \lastname{Tucker} \inst{\ref{Cardiff}}
\and  \firstname{R.} \lastname{Zylka} \inst{\ref{IRAMF}}}

\institute{\label{Cardiff} School of Physics \& Astronomy, Cardiff University, Cardiff CF24 3AA, UK  
\and \label{LLR} LLR (Laboratoire Leprince-Ringuet), CNRS, \'Ecole Polytechnique, Institut Polytechnique de Paris, Palaiseau, France
\and \label{CEFCA} Centro de Estudios de F\'isica del Cosmos de Arag\'on (CEFCA), Plaza San Juan, 1, planta 2, E-44001, Teruel, Spain          
\and \label{CEA1} AIM, CEA, CNRS, Universit\'e Paris-Saclay, Universit\'e Paris Diderot, Sorbonne Paris Cit\'e, 91191 Gif-sur-Yvette, France     
\and \label{IPAG} Univ. Grenoble Alpes, CNRS, IPAG, 38000 Grenoble, France     
\and \label{IAS} Institut d'Astrophysique Spatiale (IAS), CNRS and Universit\'e Paris Sud, Orsay, France    
\and \label{Neel} Institut N\'eel, CNRS and Universit\'e Grenoble Alpes, France
\and \label{LPSC} Univ. Grenoble Alpes, CNRS, Grenoble INP, LPSC-IN2P3, 53, avenue des Martyrs, 38000 Grenoble, France 
\and \label{Roma} Dipartimento di Fisica, Sapienza Universit\`a di Roma, Piazzale Aldo Moro 5, I-00185 Roma, Italy       
\and \label{IRAMF} Institut de RadioAstronomie Millim\'etrique (IRAM), Grenoble, France 
\and \label{CAB} Centro de Astrobiolog\'ia (CSIC-INTA), Torrej\'on de Ardoz, 28850 Madrid, Spain
\and \label{IRAME} Instituto de Radioastronom\'ia Milim\'etrica (IRAM), Granada, Spain 
\and \label{LAM} Aix Marseille Univ, CNRS, CNES, Laboratoire d'Astrophysique de Marseille, Marseille, France
\and \label{LERMA} LERMA, Observatoire de Paris, PSL Research University, CNRS, Sorbonne Universit\'es, UPMC Univ. Paris 06, 75014 Paris,
France
\and \label{Arizona} School of Earth and Space Exploration and Department of Physics, Arizona State University, Tempe, AZ 85287         
\and \label{IAP} Institut d'Astrophysique de Paris, CNRS (UMR7095), 98 bis boulevard Arago, 75014 Paris, France
\and \label{MIT} Kavli Institute for Astrophysics and Space Research, Massachusetts Institute of Technology, Cambridge, MA 02139, USA
\and \label{IRAP} Univ. Toulouse, CNRS, IRAP, 9 Av. du Colonel Roche, BP 44346, 31028, Toulouse, France
          }

\abstract{%
Understanding where and when the mass of stars is determined is one of the fundamental, mostly unsolved, questions in astronomy. Here, we present the first results of GASTON, the Galactic Star Formation with NIKA2 large programme on the IRAM 30m telescope, that aims to identify new populations of low-brightness sources to tackle the question of stellar mass determination across all masses. In this paper, we focus on the high-mass star formation part of the project, for which we map a $\sim2$ deg$^2$ region of the Galactic plane around $l=24^\circ$ in both 1.2~mm and 2.0 ~mm continuum. Half-way through the project, we reach a sensitivity of 3.7~mJy/beam at 1.2mm. Even though larger than our target sensitivity of 2~mJy, the current sensitivity already allows the identification of a new population of cold, compact sources that remained undetected in any (sub-)mm Galactic plane survey so far. In fact, about 25\%  of the $\sim 1600$ compact sources identified in the 1.2mm  GASTON image are new detections. We present a preliminary analysis of the physical properties of the GASTON sources as a function of their evolutionary stage, arguing for a potential evolution of the mass distribution of these sources with time.
}
\maketitle
\vspace{-5mm}
\section{Introduction} \label{intro}

The characteristic shape of the initial mass function (IMF) of stars has its origin rooted in the mechanisms leading to their formation. It is therefore natural to envisage the existence of different star formation regimes for every part of the IMF where the power-law slope changes. The nearly systematic association of solar-mass prestellar cores with dense, self-gravitating filaments \cite{arzoumanian2019,konyves2015} led to a picture where the peak of the IMF would be the result of the gravitational fragmentation of such filaments into individual cores, providing a compact and fixed mass reservoir for the formation of low-mass stars  inside them \cite{andre2010,andre2014}.  For high-mass stars, however, it is well known that thermal Jeans-type fragmentation cannot explain the formation of cores  more massive than a few solar masses. Hence, the formation of massive stars requires additional physics. There are strong indications that the dynamical evolution of the parsec-scale mass reservoirs surrounding massive young stellar objects play a key role. Even though the question about how massive stars form is still very much debated, observations and simulations converge towards a picture where massive stars form at the centre of globally collapsing clouds, quickly growing in mass as a result of the large infall rates  \cite[e.g.][]{bonnell2004, wang2010, peretto2013, duartecabral2013}. In this picture, massive stars are said  to be {\it clump-fed}, by opposition to low-mass star formation where stars are {\it core-fed}. To understand how the IMF is determined across all stellar masses, one now needs to focus on two additional mass intervals: Intermediate mass stars ($2$ M$_{\odot}<m_*<10$ M$_{\odot}$) for which both core-fed and clump-fed scenarios might be at work; and brown dwarfs ($m_*<0.1$ M$_{\odot}$) for which additional mechanisms such as dynamical instabilities of multiple stellar systems \cite{bate2002} or disc fragmentation \cite{thies2010} might play a role in their formation process.  
Unfortunately, no current (sub-)mm survey is either big nor sensitive enough to detect significant populations of their corresponding progenitors. 

The NIKA2 \cite{NIKA2-Adam, NIKA2-Calvo, NIKA2-Bourrion} GASTON large programme has been designed to fill that gap, with three key objectives. In addition to tackling the question of the dominant formation mechanism of brown-dwarfs in nearby clouds and mapping out a fraction of the Galactic plane at a mass sensitivity and angular resolution never reached before in a multiple square degree region, GASTON will, as a result of the dual frequency capability of NIKA2, characterise dust properties across a wide range of environments. Here, we present the first results of the Galactic plane part of the project, roughly half-way through the observation campaign. 

\begin{figure}[h]
\begin{center}
\includegraphics[scale=0.78]{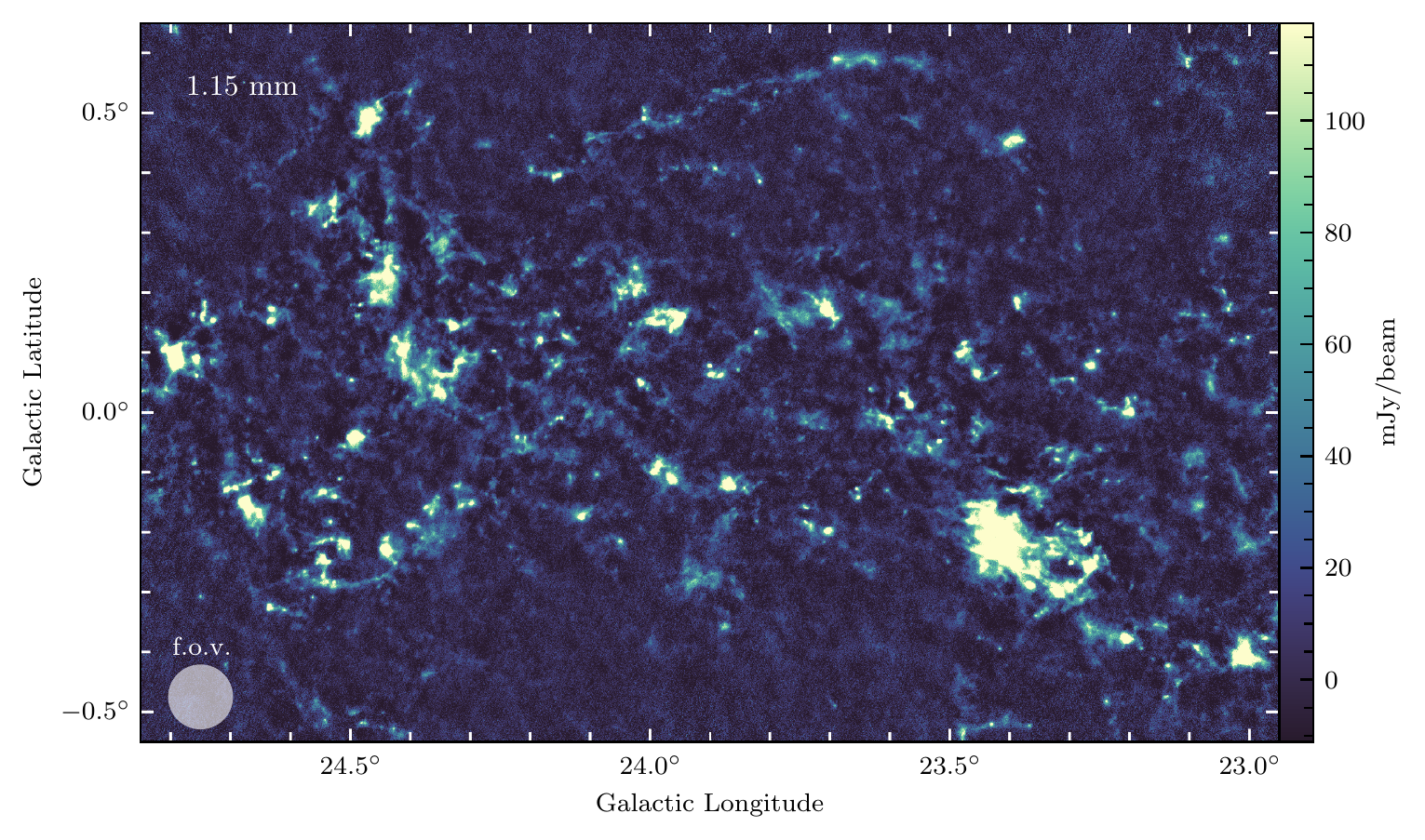}
\vspace{-5mm}
\caption{GASTON 1.2~mm map of the $\ell=24^\circ$ region of the Galactic plane, cropped to show the approximate extent of the region that should reach the target sensitivity. The grey filled circle in the bottom left corner of the image matches the NIKA2 field-of-view size.}
\vspace{-10mm}
\label{fig-1}
\end{center}
\end{figure}

\section{Observations} \label{obs}

The section of the Galactic plane targeted within GASTON is $\ell=[22.9^\circ, 24.9^\circ]$ and $b=[-0.6^\circ, +0.7^\circ]$, which has been selected for the high density of far-infrared sources identified by Hi-GAL \citep{molinari2016}. The region is observed with a series of adjoining 0.5$^\circ$ by 1.5$^\circ$ rectangular fields that cross the mid-plane at a $\pm 30^\circ$ angle. Integrations are repeated over as many times as necessary to achieve an approximately uniform sensitivity over the target area. A total area of $\sim 2.8$ deg$^{2}$ is covered, the inner $\sim 1.8~$deg$^2$ of which will reach a uniform sensitivity.  At the time of writing, about half of the observations have been done (i.e. $\sim 35$h), and the measured sensitivities are $\sigma_{1.2mm}\sim3.7$ ~mJy/beam at 1.2 mm and $\sigma_{2.0mm}\sim0.7$ ~mJy/beam at 2.0~mm. The angular resolutions are 12" and 18" at 1.2~mm and 2.0~mm, respectively. Flux calibration is regularly performed on primary calibrators such as Uranus. Pointing errors are within 3". The data has been reduced with the NIKA2 consortium IDL pipeline. The resulting 1.2~mm map is shown in Fig.~\ref{fig-1}.

\section{Compact source identification and cross-matching}

As it can be seen in Fig.~\ref{fig-1}, structures with sizes of the order of the $\sim 6.5'$  field-of-view are being recovered by the IDL reduction pipeline. In order to identify compact sources, we filter out all structures larger than an arcminute  so that  every source on the field is identified in a background-free environment. To do this, we first convolve our GASTON 1.2mm image with a Gaussian kernel to achieve an effective resolution of FWHM=60", and then subtract that convolved image from the original GASTON image, effectively filtering all sources larger than 60". We then run a dendrogram analysis \cite{rosolowsky2008} on the filtered image to identify individual sources above $3\sigma_{1.2mm}$ and compute their peak and integrated flux densities.

\begin{figure}[h]
\hspace{0.5cm}
\includegraphics[scale=0.72]{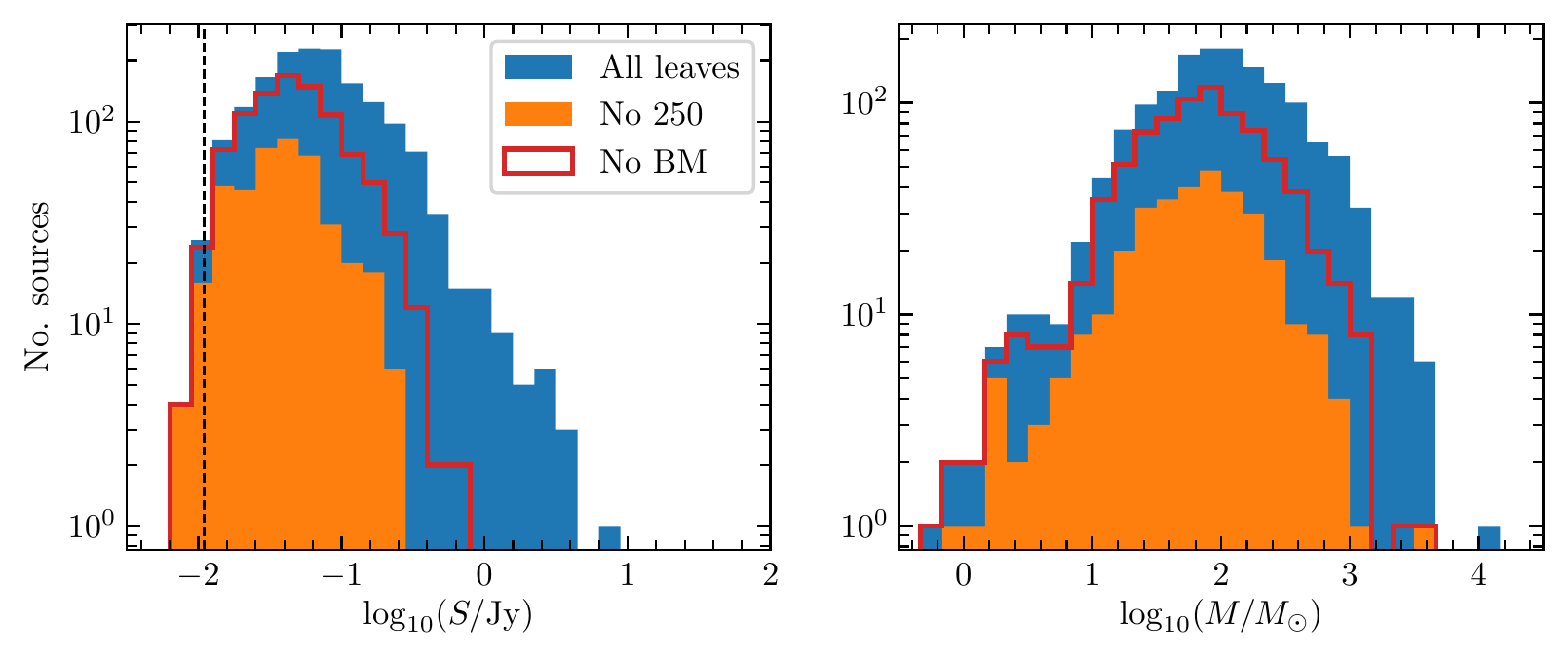}
\vspace{-5mm}
\caption{GASTON 1.2~mm compact source flux (left) and mass (right) distributions. The blue histogram shows all GASTON sources identified. The red histogram shows GASTON sources that are {\it not} associated with a band merged (BM) {\it Herschel} clump \cite{elia2017}. The orange histogram shows GASTON sources that are {\it not} associated with a {\it Herschel}  250\,$\mu$m sources from \cite{molinari2016} Hi-GAL catalogue. The vertical dashed line on the  left panel shows the 3$\sigma_{1.2mm}$ detection threshold for beam-size sources.}
\vspace{-5mm}
\label{fig-2} 
\end{figure}

In total, we have identified slightly over 1600 1.2~mm compact sources. The integrated flux distribution of these sources is shown with the blue histogram in Fig.~\ref{fig-2}-left. We can see that the histogram peaks at about 50~mJy, goes down to $\sim 10$ mJy, and up to nearly 10~Jy for the brightest source. In order to evaluate how many of these GASTON sources are already known, we cross-matched their positions with catalogued {\it Herschel} sources, the most complete, and sensitive, Galactic plane survey of cold dust structures available. For this, we used two {\it Herschel} catalogues: the band merged catalogue (BM) from \cite{elia2017}, and the 250\,$\mu$m catalogue of \cite{molinari2016}. By definition, the BM catalogue is a subsample of the 250\,$\mu$m catalogue. The histograms of GASTON sources that do not have any counterpart in these two catalogues are also displayed in Fig.~\ref{fig-2}, in which one can see that there are about 400 GASTON sources ($\sim 25\%$) that do not appear in either catalogue. These sources are, as expected, low-brightness sources (nearly all having a flux $\le 0.2$~Jy) that were most likely not picked up in {\it Herschel} images as a result of the lower far-infrared contrast of cold sources with cirrus noise. This sample of previously undetected sources opens up a brand new window on the characterisation of the earliest stages of star formation.

\section{Source properties: Distance, dust temperature, and masses}

One fundamental property of the GASTON sources is their mass. To estimate a mass from dust continuum observations, one needs three quantities: flux, distance, and dust temperature.  Fluxes of GASTON sources have been already presented in the previous section. In order to get a distance estimate for each GASTON source we used  the Galactic Ring Survey $^{13}$CO(1--0) data \cite{romanduval2010} along with the Galactic rotation model from \cite{reid2016}. For each source we get the corresponding  $^{13}$CO(1--0) spectrum averaged over the footprint of the GASTON source and, in case of multiple $^{13}$CO(1--0) velocity components, we select the strongest one as the GASTON source velocity. We then use the Bayesian distance estimator from \cite{reid2016}  to determine the distance of maximum likelihood. Finally, via a friend-of-friend algorithm, we group nearby sources with similar velocities together, and use the median distance of these sources as the single distance for that group. The idea behind this is to mitigate the effects of peculiar velocities of clumps within clusters being interpreted as differences in the distance.


Regarding the dust temperature of each GASTON source we used the same method as developed by \cite{peretto2016}, and successfully used in combination with NIKA data in \cite{rigby2018}. This method uses the ratio of the {\it Herschel} 160~$\mu$m flux to the 250~$\mu$m flux as a proxy for temperature. Here, we further filter both images using the same kernel as on the GASTON images to remove any large-scale contaminating background emission. From the {\it Herschel} filtered images one can get a dust temperature at 18" resolution (the resolution of the 250~$\mu$m image) for every GASTON source.  With distances, and dust temperatures one can  now compute the mass distributions of the GASTON compact sources. For this we assumed the following specific dust opacity law $\kappa_\nu = 0.1 \, ( \nu / 1 \, \mathrm{THz})^{1.8} \, \mathrm{cm}^2 \, \mathrm{g}^{-1}$, which accounts for a 1\% dust-to-gas mass ratio \cite{andre2010,rigby2018}.  The mass distributions of the three samples (see previous section) are shown in Fig.~\ref{fig-2}-right. The striking feature of this figure is the fact that  sources that were not identified before in any other catalogue (orange histogram) have a very similar mass distribution to rest of the sample, despite having lower fluxes. This shows that this new sample is mostly made of cold compact and relatively massive sources that were just not picked up by any other survey. This source population could represent an important missing building block in our understanding of the stellar mass determination. 

\section{Evolution of mass distributions}

\begin{figure}[t]
\hspace{0.5cm}
\includegraphics[scale=0.68]{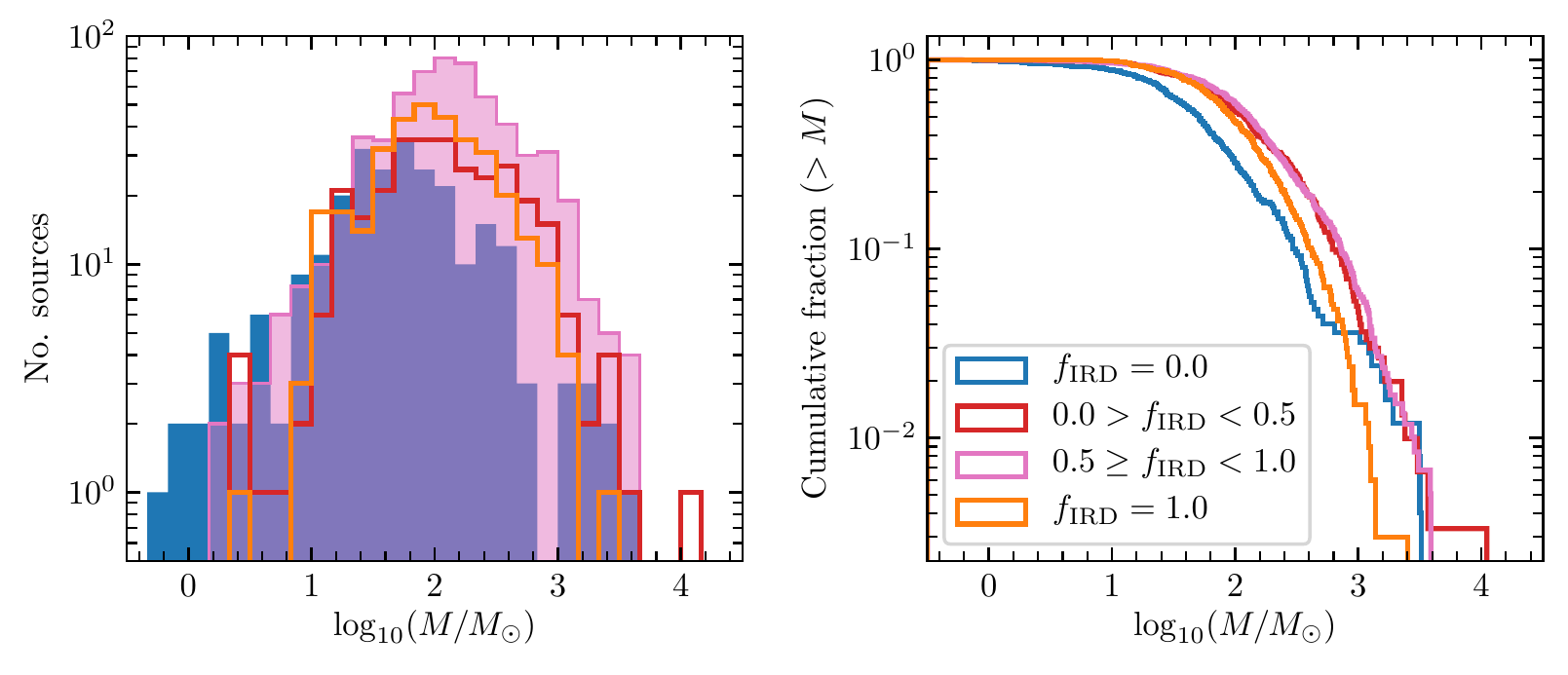}
\vspace{-3mm}
\caption{GASTON 1.2~mm compact source mass distributions as a function of their infrared darkness parameter. The histograms show the mass distributions of sources with: i) $f_\mathrm{IRD} = 1$ (i.e. completely infrared-dark) in orange; ii)  $1 > f_\mathrm{IRD} > 0.5$ in pink; iii) $0.5 > f_\mathrm{IRD} > 0$ in blue and iv) $f_\mathrm{IRD} = 0$ (i.e. completely infrared-bright) in blue.  The differential distributions are displayed on the  left-hand-side panel, the cumulative forms are displayed in the right-hand-side panel. }
\vspace{-5mm}
\label{fig-4}  
\end{figure}

As dusty, star-forming clumps evolve, embedded young-stars heat up an increasing fraction of the clump dust mass. During that process, a cold clump that was originally dark in the mid-infrared can become infrared bright. One can therefore use the infrared darkness of a clump as a proxy for its evolutionary stage \cite{battersby2017}. Using a combination of {\it Spitzer} 8~$\mu$m and {\it Herschel} data, Watkins, Peretto et al. (in prep) have developed a method to compute the fraction of a clump mass that is infrared dark. We here use that method to characterise the infrared darkness of GASTON sources. We then split our  sample  into four bins of infrared darkness parameters, from $f_{IRD}=1$ (totally infrared dark) to  $f_{IRD}=0$ (totally infrared bright), and we plot the mass distributions for each of the four subsamples. This is shown in Fig.~\ref{fig-4}. This figure tentatively shows that there are differences in the shape of the mass distribution, with the most massive sources having intermediate values of infrared darkness parameter. If one considers infrared darkness $f_{IRD}$ as a valid proxy for time evolution, then this would suggest that sources gain mass as they evolve from $f_{IRD}=1$ to $f_{IRD}=0.5$, and then lose mass as they evolve towards $f_{IRD}=0$. This is what one expects in a scenario where core and protostellar mass growth occur simultaneously \cite{hatchell2008}. 

\section{Summary}

We used NIKA2 on the IRAM 30m telescope to map over 2 square degrees of the Galactic plane at 1.2~mm and 2.0~mm. While the observing campaign is still ongoing we performed a first analysis of the 1.2~mm image. The high sensitivity of these GASTON images allowed us to identify a brand new population of low-brightness compact sources, which, being amongst the coldest sources, have a similar mass distribution as those which were already known. This new sample might well represent a missing building-block in our understanding of the stellar mass determination process. Using infrared darkness as a possible proxy of their evolutionary stage, we also suggest that there might be some evidence for a time evolution of the clump mass distribution. However, further detailed analysis of the data is required to evaluate the robustness of such interpretation. The remaining integration time of the GASTON field will allow us to increase source statistics on cold intermediate-mass sources ($\sim 10$~M$_{\odot}$), allowing the connection of our results with those obtained in well-known nearby star-forming clouds. In this context, the GASTON 2.0~mm data will be instrumental in confirming the nature of low signal-to-noise ratio sources detected at 1.2~mm.

\section*{Acknowledgements}
 N. P. and  A. R. would like to thank STFC for support under the consolidated grant number ST/N000706/1. We thank the Royal Society for providing computing resources under Research Grant number RG150741. We would also like to thank the IRAM staff for their support during the campaigns. The NIKA dilution cryostat has been designed and built at the Institut N\'eel. In particular, we acknowledge the crucial contribution of the Cryogenics Group, and in particular Gregory Garde, Henri Rodenas, Jean Paul Leggeri, Philippe Camus. This work has been partially funded by the Foundation Nanoscience Grenoble and the LabEx FOCUS ANR-11-LABX-0013. This work is supported by the French National Research Agency under the contracts "MKIDS", "NIKA" and ANR-15-CE31-0017 and in the framework of the "Investissements d’avenir” program (ANR-15-IDEX-02). This work has benefited from the support of the European Research Council Advanced Grant ORISTARS under the European Union's Seventh Framework Programme (Grant Agreement no. 291294). We acknowledge fundings from the ENIGMASS French LabEx (R. A. and F. R.), the CNES post-doctoral fellowship program (R. A.), the CNES doctoral fellowship program (A. R.) and the FOCUS French LabEx doctoral fellowship program (A. R.). R.A. acknowledges support from Spanish Ministerio de Econom\'ia and Competitividad (MINECO) through grant number AYA2015-66211-C2-2.

%

\begin{thebibliography}{}
%
%

%


\newcommand{\apj}{Astrophys. J.}
\newcommand{\apjs}{Astrophys. J. Supp.}
\newcommand{\aanda}{Astron. Astrophys.}
\newcommand{\mnras}{Mon. Notices Royal Astron. Soc.}


\bibitem{arzoumanian2019}
D.~Arzoumanian {\it et al.}, \aanda. {\bf 621}, 42 (2019)

\bibitem{konyves2015}
V.~K\"onyves {\it et al.}, \aanda. {\bf 584}, A91 (2015)


\bibitem{andre2010}
P.~Andr\'{e} {\it et al.}, \aanda. {\bf 518}, 102 (2010)

\bibitem{andre2014}
P.~Andr\'{e} {\it et al.}, \textit{Protostars and Planets VI}. Univ. of Arizona Press, Tucson, p.27 (2014)





\bibitem{peretto2013} 
N.~Peretto  {\it et al.}, \aanda. {\bf 555}, A112 (2013)


\bibitem{duartecabral2013}
A.~Duarte-Cabral {\it et al.}, \aanda. {\bf 558}, A125 (2013)


\bibitem{bonnell2004}
I.~A.~Bonnell {\it et al.}, \mnras. {\bf 349}, 735 (2004)


\bibitem{wang2010}
P.~Wang {\it et al.}, \apj. {\bf 709}, 27 (2010)

\bibitem{bate2002}
M.~Bate {\it et al.}, \mnras. {\bf 332}, 65 (2002)

\bibitem{thies2010}
I.~Thies {\it et al.}, \apj. {\bf 717}, 577 (2010)


\bibitem{NIKA2-Adam}
R.~Adam {\it et al.}, \aanda.\  {\bf 609}, A115 (2018)

\bibitem{NIKA2-Calvo}
M. Calvo {\it et al.}, Journal of Low Temperature Physics, {\bf 184}, 816 (2016)

\bibitem{NIKA2-Bourrion}
O.~Bourrion {\it et al.}, Journal of Instrumentation. {\bf 11}, P11011 (2016)


\bibitem{molinari2016}
S.~Molinari {\it et al.}, \aanda. {\bf 591}, A149 (2016)



\bibitem{rosolowsky2008}
E.~Rosolowsky {\it et al.}, \apj. {\bf 679}, 1338--1351, (2008)

\bibitem{elia2017}
D.~Elia {\it et al.}, \mnras {\bf 471}, 100--143 (2017)

\bibitem{romanduval2010}
J.~Roman-Duval {\it et al.}, \apj, {\bf 723}, 492


\bibitem{reid2016}
M.~J.~Reid {\it et al.}, \apj, {\bf 823}, 77 (2016)

\bibitem{peretto2016}
N.~Peretto {\it et al.}, \aanda, {\bf 590}, A72 (2016)

\bibitem{rigby2018}
A.~Rigby {\it et al.}, \aanda, {\bf 615}, A18 (2018)

\bibitem{battersby2017}
C.~Battersby {\it et al.}, \apj. {\bf 835}, 263 (2017)

\bibitem{hatchell2008}
J.~Hatchell \& G.~Fuller \aanda {\bf 482}, 855--863 (2008)


%





%
%


\end{thebibliography}
%
%

\end{document}